\documentclass[12pt]{article}
\usepackage{a4}
\usepackage{epsfig}
\usepackage[pagewise]{lineno}

\newcommand{\btaunux}
{\ifmmode   \protect{{\rm b}\to\tau^{-}\overline{\nu}_\tau {\rm X}}
 \else     \protect{${\rm b}\to\tau^{-}\overline{\nu}_\tau {\rm X}$ }\fi}
\newcommand{\benux}
{\ifmmode   \protect{{\rm b}\to e^{-}\overline{\nu}_{\rm e} {\rm X}}
 \else     \protect{${\rm b}\toe^{-}\overline{\nu}_{\rm e} {\rm X}$ }\fi}
\newcommand{\blnux}{b $\rightarrow \ell^- \overline{\nu} {\rm X}$}
\newcommand{\bclnux}{b $\rightarrow$ c or $\overline{c}\rightarrow \ell \nu {\rm X}$ }

\newcommand{\Z}{${\rm Z}$ }

\newcommand{\Ztautau}{${\rm Z} \rightarrow \tau^+ \tau^- $ }

\newcommand{\evis}{E_{\rm vis}^{\rm hemi}}
\newcommand{\emis}{E_{\rm miss}^{\rm hemi}}
\newcommand{\ebeam}{E_{\rm beam}}
\newcommand{\ecorr}{E_{\rm corr}}
\newcommand{\ee}{\ifmmode {\rm e}^+{\rm e}^- \else e$^+$e$^-$\fi}
\newcommand{\qqbar}{\ifmmode {\rm q}\bar{\rm q} \else q$\bar{\mbox{\rm q}}$\fi}
\newcommand{\ccbar}{\ifmmode {\rm c}\bar{\rm c} \else c$\bar{\mbox{\rm c}}$\fi}
\newcommand{\bbbar}{\ifmmode {\rm b}\bar{\rm b} \else b$\bar{\mbox{\rm b}}$\fi}
\newcommand{\dedx}{\ifmmode {\rm d}E/{\rm d}x \else d$E$/d$x$\fi}
\newcommand{\Dzero}{\ifmmode {\rm D}^0 \else D$^0$\fi}
\newcommand{\Dstar}{\ifmmode {\rm D}^* \else D$^*$\fi}
\newcommand{\Dstarp}{\ifmmode {\rm D}^{*+} \else D$^{*+}$\fi}
\newcommand{\Dss}{\ifmmode {\rm D}^{**} \else D$^{**}$\fi}
\newcommand{\dst}{\ifmmode {\rm D}_{\rm s}^{-} \else D$_{\rm s}^{-}$\fi}
\newcommand{\dstaunu}{\mbox{D$_{\rm s}^- \rightarrow \tau^- \overline{\nu}_\tau$}}

\newcommand{\ntot}{$3.70$ }
\newcommand{\brbtau}{2.78}
\newcommand{\brbtaustat}{0.18}
\newcommand{\brbtausyst}{0.51}

\newcommand{\tanblim}{0.53}
\newcommand{\nsel}{$90587$ }
\begin{document}
\begin{titlepage}
\begin{center}{\large  EUROPEAN ORGANIZATION FOR NUCLEAR RESEARCH
}\end{center}\bigskip
\begin{flushright}
       CERN-EP-2001-058   \\ 31 July 2001
\end{flushright}
\bigskip\bigskip\bigskip\bigskip\bigskip
\begin{center}{\huge\bf 
\mbox{ \hskip -1cm Measurement of the Branching Ratio}
\mbox{for the Process \boldmath$\btaunux$\unboldmath}
}\end{center}\bigskip\bigskip
\begin{center}{\LARGE The OPAL Collaboration
}\end{center}\bigskip\bigskip
\bigskip\begin{center}{\large  Abstract}\end{center}
The inclusive branching ratio for the process \btaunux
has been measured using hadronic \Z decays collected by the OPAL experiment
at LEP in the years 1992-2000. The result is:
$$\mathrm{BR}(\btaunux) = (\brbtau \pm \brbtaustat \pm \brbtausyst) \%$$
This measurement is consistent with the Standard Model expectation and 
puts a constraint of 
$$\tan\beta / M_{\rm H}^{\pm} < \tanblim\ {\rm GeV}^{-1}$$
at the $95\%$ confidence level on Type II Two Higgs Doublet Models.
\bigskip\bigskip

\bigskip\bigskip
\begin{center}{\large
(Submitted to Physics Letters B)
}\end{center}
\end{titlepage}

\begin{center}{\Large        The OPAL Collaboration
}\end{center}\bigskip
\begin{center}{
G.\thinspace Abbiendi$^{  2}$,
C.\thinspace Ainsley$^{  5}$,
P.F.\thinspace {\AA}kesson$^{  3}$,
G.\thinspace Alexander$^{ 22}$,
J.\thinspace Allison$^{ 16}$,
G.\thinspace Anagnostou$^{  1}$,
K.J.\thinspace Anderson$^{  9}$,
S.\thinspace Arcelli$^{ 17}$,
S.\thinspace Asai$^{ 23}$,
D.\thinspace Axen$^{ 27}$
% ...
}\end{center}\bigskip
\bigskip
$^{  1}$School of Physics and Astronomy, University of Birmingham,
Birmingham B15 2TT, UK
\newline
$^{  2}$Dipartimento di Fisica dell' Universit\`a di Bologna and INFN,
I-40126 Bologna, Italy
\newline
$^{  3}$Physikalisches Institut, Universit\"at Bonn,
D-53115 Bonn, Germany
\newline
% ...
$^{  5}$Cavendish Laboratory, Cambridge CB3 0HE, UK
\newline
% ...
$^{  9}$Enrico Fermi Institute and Department of Physics,
University of Chicago, Chicago IL 60637, USA
\newline
% ...
$^{ 16}$Department of Physics, Schuster Laboratory, The University,
Manchester M13 9PL, UK
\newline
$^{ 17}$Department of Physics, University of Maryland,
College Park, MD 20742, USA
\newline
% ...
$^{ 22}$Department of Physics and Astronomy, Tel Aviv University,
Tel Aviv 69978, Israel
\newline
$^{ 23}$International Centre for Elementary Particle Physics and
Department of Physics, University of Tokyo, Tokyo 113-0033, and
Kobe University, Kobe 657-8501, Japan
\newline
% ...
$^{ 27}$University of British Columbia, Department of Physics,
Vancouver BC V6T 1Z1, Canada
\newline
% ...
Tel Aviv 69978, Israel.

\section{Introduction}
\label{sec:intro}
This paper describes a measurement of the inclusive branching ratio 
BR(\btaunux), using data taken with the OPAL detector at LEP in 
the years 1992-2000 at e$^+$e$^-$ centre-of-mass energies around the 
Z resonance. 
Similar measurements of BR(\btaunux) have been published previously by
the other LEP experiments~[1-3].
The measurements can be directly compared to the Standard 
Model expectation calculated
in the framework of Heavy Quark Effective Theory (HQET)~\cite{bib:HQET},
and so can be used to constrain basic parameters of
HQET~\cite{bib:HQETPAR}.

A measurement of BR($\btaunux$) is also a probe for the 
presence of a new charged boson coupling to mass.
This coupling would increase the branching ratio 
BR($\btaunux$)~\cite{bib:beyond,bib:gross}. 
Since a charged Higgs boson
contributes at tree level, its contribution cannot
be easily cancelled by other new particles.
This can be used to set a limit on a contribution of the 
charged Higgs exchange.
In the Minimal Supersymmetric 
Standard Model (MSSM), however, a region of the parameter space
is found where one-loop SUSY-QCD corrections could 
weaken the bound~\cite{bib:susyqcd}.

Many extensions of the Standard Model, like the MSSM, 
include Type II Two Higgs Doublets, 
where one of two Higgs doublets couples
only to down-type quarks and the other only to up-type quarks. 
In these models, $\tan\beta$ is the ratio of the vacuum expectation
values of the two Higgs doublets and $M_{\rm H}^{\pm}$ is the mass of the
charged Higgs boson. 
The decay rate for \btaunux\ can be calculated
as a function of $r=\tan\beta/M_{\rm H}^{\pm}$.
A term proportional to $r^2$ is added to the Standard Model decay rate
of ${\rm BR}(\btaunux)=(2.36 \pm 0.17)\%$.
A value of $r=0.5$~GeV$^{-1}$, for example, yields
BR($\btaunux)=(3.61\pm0.36)\%$~\cite{bib:gross}. 
The errors are fully correlated and include the 
experimental uncertainty on 
${\rm BR}(\benux)=(10.86\pm0.35)\%$~\cite{bib:pdg00} and the theoretical 
uncertainties.

Since direct reconstruction of the $\tau$ lepton in a multihadronic event
is not possible, other properties of the signal decays have to be 
exploited. Each event is divided into two hemispheres.
Hemispheres containing \btaunux decays are characterised 
by large missing energy, due to the presence of at least two neutrinos in the 
final state. The reconstruction of the missing energy uses
the OPAL calorimeters and tracking detectors and it relies on
the hermeticity of the detector.
To select a sample enriched in b decays, 
a b tagging algorithm is applied. 
The b tagging is done in the hemisphere opposite to the signal
to reduce the dependence on the Monte Carlo simulation of the signal.

Semileptonic b decays like \blnux~($\ell={\rm e},\mu$)
are an important background.
They are suppressed by rejecting hemispheres with an identified 
electron or muon. The same lepton veto suppresses the leptonic decays of 
the $\tau$ lepton in signal events, thus selecting mostly hadronic 
$\tau$ decays. 

\section{Detector, data set and Monte Carlo samples}
\label{sec:det}

The details of the 
construction and performance of the OPAL detector are described 
elsewhere~\cite{bib:opal}. Here only the main components relevant 
for this analysis are described.

Tracking of charged particles is performed by a central detector,
consisting of a silicon microvertex detector, a vertex chamber, a jet chamber
and $z$-chambers\,\footnote{A right handed coordinate system is used, with
positive $z$ along the $\mathrm{e}^-$ beam direction and $x$ pointing
towards the centre of the LEP ring. The polar and azimuthal angles are
denoted by $\theta$ and $\phi$, and 
the origin is taken to be the centre of the detector.}.
The central detector is inside a
solenoid, which provides a uniform axial magnetic field of 0.435\,T.
The silicon microvertex detector consists of two layers of
silicon strip detectors; for most of the data used in this paper,
the inner layer covered a polar angle range of $|\cos\theta |<0.83$ and
the outer layer covered $| \cos \theta |< 0.77$, with
an extended coverage for the data taken after the year 1996.
The vertex chamber is a precision drift chamber
which covers the range $|\cos \theta | < 0.95$.
The jet chamber is
a large-volume drift chamber, 4.0~m long and 3.7~m in diameter,
providing both tracking and ionisation energy loss (d$E$/d$x$) information.
The $z$-chambers provide a precise measurement of the $z$-coordinate
of tracks as they leave the jet chamber in the range
$|\cos \theta | < 0.72$.

Immediately outside the tracking volume is the solenoid and a time-of-flight 
counter array followed by an electromagnetic shower presampler and
a lead-glass electromagnetic calorimeter. 
The return yoke of the magnet lies outside the electromagnetic
calorimeter and is instrumented with limited streamer chambers.
It is used as a hadron calorimeter and assists in the reconstruction
of muons. The outermost part of the detector is made up by
layers of muon chambers.

Hadronic Z decays collected with the OPAL detector
at e$^+$e$^-$ centre-of-mass energies around the 
Z resonance are selected using a standard multihadron 
selection~\cite{bib:TKMH}. 
To reduce further the small contribution of \Ztautau decays
an additional requirement of at least 7 tracks in each event is imposed.
With these criteria, the selection efficiency for hadronic Z decays
is $(98.1 \pm 0.5)\%$~\cite{bib:gambb} with a background of $(0.11\pm 0.03)$\%.
After hadronic event selection and \Ztautau rejection, the resulting data 
sample collected 
in the years 1992-2000 after the installation of 
the silicon microvertex detector consists of $\ntot\times10^6$ events.
About $11\%$ of the data used were recorded in the years 1996-2000,
for calibration purposes.

A Monte Carlo sample of hadronic Z decays of about seven times
the size of the recorded data sample for b flavour events and about 
the same size as the recorded data for the other flavours is used in the 
analysis.
The Monte Carlo events are generated using JETSET~7.4~\cite{bib:jetset} 
with the b and c quark fragmentation modelled according to the 
parameterisation of Peterson et al.~\cite{bib:peter}.
A global fit to OPAL 
data has been performed to optimise the JETSET parameters~\cite{bib:tuned}. 
The decay \btaunux is modelled in JETSET using matrix elements neglecting the
mass of the final state particles.
The energy distribution of the $\tau$ lepton in the b rest frame is therefore
reweighted to reproduce the energy distribution calculated by including mass 
effects~\cite{bib:HQET,bib:etau}.
The polarisation of the $\tau$ leptons is simulated by interfacing JETSET 
to the TAUOLA~\cite{bib:tauola} $\tau$ decay simulation package. 
The $\tau$ polarisation is calculated according to~\cite{bib:HQET} 
with the HQET parameters $\lambda_1$ and $\lambda_2$ set to zero, 
corresponding to the free quark model. 

All events have been processed using a full simulation of the 
OPAL detector \cite{bib:GOPAL} and the same reconstruction algorithms 
that were applied to the data.

\section{Event and hemisphere selection}
\label{sec:sel}

Each event is divided into two hemispheres using the 
thrust variable, $T$, which is defined by~\cite{bib:thrust}
\begin{equation}
T= \max_{\vec{n}}\left(\frac{\sum_i|\vec{p}_i\cdot\vec{n}|}
                   {\sum_i|\vec{p}_i|}\right),
\label{equ_thrust}
\end{equation}
where $\vec{p}_i$ are the momentum vectors of the particles in an event.
The thrust axis $\vec{n}_T$ is the direction $\vec{n}$ which
maximises the expression in parenthesis.  A plane through the origin
and perpendicular to $\vec{n}_T$ divides the event into two
hemispheres.

Signal hemispheres are characterised by large missing energy.
Particles escaping close to the beam pipe can fake the large 
missing energy signature of neutrinos. 
By requiring the polar angle 
$\theta$ of the thrust axis $\vec{n}_T$ of the event
to satisfy $|\cos{\theta}| < 0.8$, only events well contained in 
the central detector are accepted.
Events with a two-jet topology are expected to
have thrust values close to one and are therefore selected by 
requiring the thrust $T>0.85$. 

The b-tagging algorithm~\cite{bib:btag} gives a likelihood 
that a hemisphere originates from a b decay by combining the information 
from a secondary vertex neural network, a jetshape neural network
and a prompt lepton finder. 
A cut on the b likelihood was applied giving a 
selection efficiency of $47\%$ and 
a purity of $92\%$ for b flavour hemispheres in the selected sample 
according to Monte Carlo.

The hemisphere opposite to the signal hemisphere is used for b-tagging.
This reduces the dependence on the Monte Carlo description for the signal
since the b tagging efficiency can be measured directly from the
data using events with one and two tagged hemispheres.

\section{Missing energy distribution}
\label{sec:emis}
In each hemisphere the missing energy $\emis$ is calculated by
\begin{equation} 
\emis = \ebeam + \ecorr - \evis. 
\end{equation}
The sum of the beam energy $\ebeam$
and the correction term $\ecorr$ is the predicted energy in the hemisphere.
The missing energy $\emis$ is obtained by subtracting the
visible energy $\evis$.
The correction term $\ecorr$ is determined
by exploiting the overall energy and momentum conservation in the Z decay. 
Assuming a decay of the Z boson into two particles,
the correction term is
\begin{equation}
\ecorr = \frac{M_{\rm hemi}^2-M_{\rm ohemi}^2}{4 \ebeam},
\label{eq:ecorr}
\end{equation}
where $M_{\rm hemi}$ and $M_{\rm ohemi}$ are the measured invariant 
masses of the signal hemisphere and of the opposite hemisphere.

The visible energy $\evis$ in the hemisphere is obtained by summing  
separately the energies of charged particles reconstructed in 
the central detector and neutral particles depositing energy in the 
electromagnetic calorimeter. 
A matching algorithm \cite{bib:MT} is used to associate tracks in the central 
detector with clusters in the electromagnetic calorimeters. This algorithm 
corrects the measured electromagnetic calorimeter cluster energy if a track 
is matched to the cluster. 
Tracks are counted only if they have been 
reconstructed using at least 20 jet chamber hits, have a $p_{\rm t}$ 
of at least 120~MeV with respect to the beam axis 
and a distance of closest approach $d_{0}$ to the 
beam axis of 
less than 2.5 cm. Electromagnetic clusters are counted if they have at least 
100~MeV energy in the barrel region or 250~MeV in the forward region 
of the detector.

Only hemispheres with missing energy $\emis > 5$~GeV are used.
The cut has been optimised to maximise
the statistical sensitivity of the fit whilst minimising the systematic
uncertainty due to the Monte Carlo 
description of the missing energy distribution. 
The optimisation is discussed in Section~\ref{sec:syst}.

To remove background from non-$\tau$ semileptonic b decays, hemispheres with 
identified electrons or muons are rejected. 
Neural networks are used to identify electrons~\cite{bib:elec1}
and muons~\cite{bib:muon} with momenta larger than 2~GeV.
In the selected Monte Carlo sample $91\%$ of the hemispheres with
prompt electrons or muons from semileptonic b and c 
decays are rejected with this requirement.
Tracks tagged as originating from photon conversions~\cite{bib:conv} 
in the tracking detector
are not accepted as electron candidates.

\begin{figure}[htbp]
  \begin{center}
    \epsfig{figure=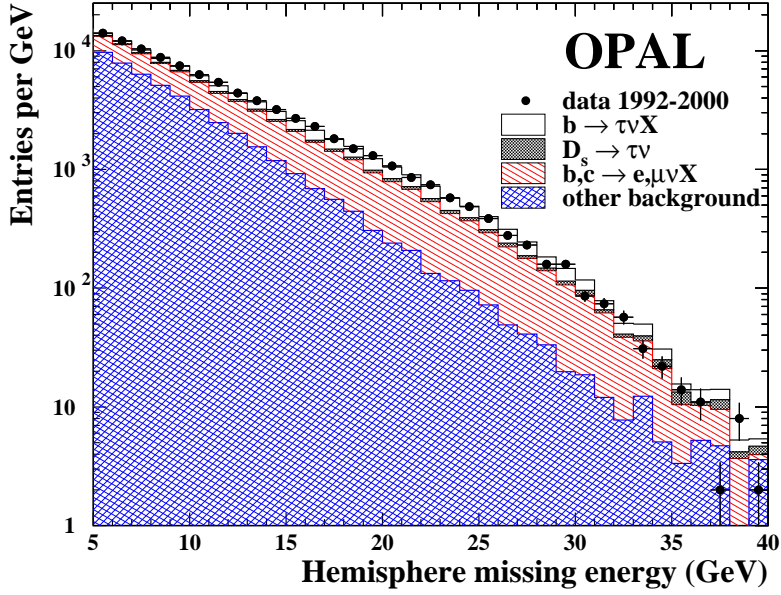,width=0.8\textwidth}
    \epsfig{figure=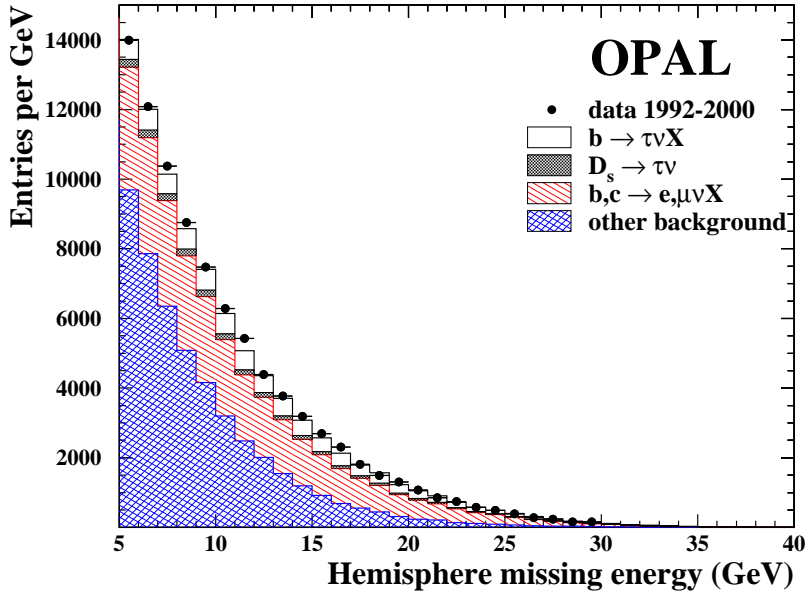,width=0.8\textwidth}
    \caption{Distribution of the missing energy in a hemisphere, $\emis$,
     for all selected data events and for the Monte Carlo simulation shown
     on a logarithmic scale and on a linear scale.
     The backgrounds from semileptonic heavy quark decays into electrons
     and muons, and from \dstaunu~decays are shown separately.
     The number of \btaunux signal events is set to the result of the fit.}
    \label{fig:result}
  \end{center}
\end{figure}
A sample of \nsel hemispheres remains in the data after applying all cuts.
Using the Monte Carlo it is estimated to contain about $9\%$ signal events, 
 $3\%$ $\dstaunu$ decays, $35\%$ non-$\tau$ semileptonic b and c decays, 
$4\%$ light quark decays and $49\%$ hadronic b and c decays.
A large fraction of the semileptonic b and c decays are decays with
leptons with momenta below $2$~GeV.
Using the $\emis$ distributions predicted by the Monte Carlo simulation
for signal and the sum of all backgrounds, 
a binned maximum likelihood fit is used to determine 
BR(\btaunux) with the fraction of signal events as the only free parameter.
The result of the fit is
\begin{equation}
\mathrm{BR}(\btaunux) =  (\brbtau \pm \brbtaustat) \%,
\end{equation}
where the uncertainty takes into account
the limited number of Monte Carlo and data events~\cite{bib:fit}. 
The result is obtained by fitting the data samples of the different
years with the corresponding Monte Carlo samples.
The minimum of the fits correspond to $\chi^2$ values for $34$~degrees
of freedom in the range $31$ to $45$. 
The $\emis$ distribution for the total data sample
is shown in Fig.~\ref{fig:result}.

\section{Systematic uncertainties}
\label{sec:syst}
The measurement of BR(\btaunux) relies on the Monte Carlo modelling 
of the missing energy distribution for the signal and background events.
Other systematic uncertainties arise from the reproduction of the selection 
and veto efficiencies for the data in the Monte Carlo simulation
and from limited knowledge of the branching ratios for decays
involving a heavy quark and a lepton.
The sources of systematic uncertainties are described below and 
their contributions to
the total systematic uncertainty are summarised in Table~\ref{tab:syst}.
\begin{table}[htbp]
  \begin{center}
    \begin{tabular}{|l|c|} 
      \hline
      Source & $\Delta$BR(\btaunux)  \\ 
      \hline
      leptonic $\emis$ description         & $-0.16\%$ \\
      hadronic $\emis$ description         & $+0.15\%$ \\
      \hline
      tracking resolution                  & $\pm0.04\%$ \\
      calorimeter resolution               & $\pm0.06\%$ \\
      b tagging efficiency                 & $\pm0.04\%$ \\
      e$^{\pm}$ veto                       & $\pm0.11\%$ \\
      $\mu^{\pm}$ veto                     & $\pm0.07\%$ \\
      \hline
      $\langle x_{\rm b} \rangle = 0.702 \pm 0.008~\cite{bib:pr319}$     
                                          & $\pm0.33\%$ \\
      $\langle x_{\rm c} \rangle = 0.484 \pm 0.008~\cite{bib:pr319}$     
                                          & $\pm0.04\%$ \\
      \btaunux decay modelling            & $\pm0.03\%$ \\
      semileptonic b decay models          & $\pm0.26\%$ \\
      BR(\blnux)=($10.73\pm0.18)\%$~\cite{bib:pdg00}            &$\pm0.08\%$ \\
      BR(\bclnux)=($9.69\pm0.51)\%$~\cite{bib:pr319}            &$\pm0.05\%$ \\
      BR(\dstaunu)=($7.2\pm2.3)\%$~\cite{bib:dstau,bib:dstaulep}&$\pm0.13\%$ \\
      $f({\rm b}\to\dst)=(18\pm5)\%)$~\cite{bib:fbs}            &$\pm0.10\%$ \\
      \hline
      Total systematic uncertainty         &$\pm \brbtausyst$\%\\
      \hline
    \end{tabular}
    \caption{The contributions to the systematic uncertainty
 on BR(\btaunux).} 
    \label{tab:syst}
  \end{center}
\end{table}

An imperfect modelling of the signal region of the $\emis$ distribution by the
Monte Carlo simulation can bias the result.
We have therefore studied the modelling of the $\emis$ distribution 
using three different signal-depleted control samples: a sample 
enriched in semileptonic b decays, a sample enriched in hadronic
b decays and a light quarks control sample.
In these samples, the ratio of the
$\emis$ distributions of data and Monte Carlo has been studied. 
The missing energy distributions for the control samples together with the
corresponding Monte Carlo predictions are shown in Fig.~\ref{fig:control}.

\begin{description}

\item[Leptonic control sample:]
A sample enriched in semileptonic b decays is selected by using 
the same requirements for the b-tagging as for the signal 
sample (Section~\ref{sec:sel}). 
For the leptonic control sample, hemispheres with electrons and muons
are selected and not rejected.
To obtain a pure sample, at least one electron or one muon candidate with 
a high value for the output of the
lepton identification neural networks is required.
According to the Monte Carlo about $85\%$ of the control sample 
are semileptonic b and c decays.

The $\emis$ distribution for this class of events is well described 
by the Monte Carlo simulation (Fig.~\ref{fig:control}a).
To estimate the systematic uncertainty associated with residual 
differences, the ratio of the data and the Monte Carlo distribution
(Fig.~\ref{fig:control}b)
is fitted with a straight line separately for every year of data taking. 
The largest slope $a$ obtained from these fits
is used to reweight the $\emis$ distribution for
semileptonic b decays in the Monte Carlo background. 
The events are reweighted using a weight 
\begin{equation}
\label{eq:w}
w(\emis)= 1 + a(\emis-\overline{E}^{\rm hemi}_{\rm miss}),
\end{equation} 
where $\overline{E}^{\rm hemi}_{\rm miss}$ 
is the mean of the missing energy distribution. 
Using a second order polynomial fit instead
yields a similar systematic uncertainty.

\item[Hadronic control sample:]
To study the description of the missing energy distribution for
hadronic b decays, a sample is selected by applying b-tagging, 
the lepton veto and by requiring at least 10~GeV energy deposit 
in the hadron calorimeter. 
These requirements enrich the sample in events with 
large hadronic energy. The agreement between data and Monte Carlo 
for the $\emis$ distribution is reasonable in the signal 
region (Fig.~\ref{fig:control}c,d).
The systematic uncertainty is estimated using the method described above 
and given in Table~\ref{tab:syst}. 
It increases if the cut on $\emis$
is reduced (Table~\ref{tab:ecut}). To keep the quadratic sum
of the statistical and systematic uncertainty small, 
the cut $\emis > 5$~GeV is chosen. All other relevant systematic uncertainties
show only a small dependence on the $\emis$ cut.

\begin{table}[htbp]
  \begin{center}
    \begin{tabular}{|c|c|c|} 
      \hline
      $\emis$ cut& stat & syst \\ 
      (GeV)      &      &      \\
      \hline
      0  & $\pm 0.14\%$ & $+1.00\%$ \\
      3  & $\pm 0.16\%$ & $+0.45\%$ \\
      5  & $\pm 0.18\%$ & $+0.15\%$ \\
      7  & $\pm 0.21\%$ & $+0.09\%$ \\
      9  & $\pm 0.27\%$ & $+0.08\%$ \\
      \hline
    \end{tabular}
    \caption{The statistical and the systematic uncertainty
    on the measurement of BR(\btaunux) due to
    the Monte Carlo description for events with hadronic b and c decays 
    for different fitting ranges.} 
    \label{tab:ecut}
  \end{center}
\end{table}

\item[Light quarks control sample:]
A further test of the Monte Carlo description of the $\emis$ distribution
is performed for a light quark (uds) flavour sample. 
The sample is selected by exploiting the fact that
events with a hadron carrying reconstructed 
momentum between 0.5 and 1.07 of the beam 
momentum mostly originate from uds primary quarks~\cite{bib:highx}. 
The light quark tag is applied in the opposite hemisphere.
The $\emis$ distribution for this sample 
is shown in Fig.~\ref{fig:control}e. 
Reweighting the light quark background events according to Eq.~\ref{eq:w} 
results in a negligible contribution to the total systematic uncertainty.
\item[Detector effects:]
The description of the visible energy distribution depends on a correct 
reproduction of the 
detector resolution in the Monte Carlo samples. The tracking resolution has 
been modified by $\pm 10\%$ to evaluate the systematic uncertainty 
\cite{bib:elec1}. The 
energy scale of the electromagnetic calorimeter (ECAL) is varied 
according to the small shift between the mean of the 
ECAL energy distributions in data and Monte Carlo found for 
an inclusive sample of hadronic Z decays. 
The uncertainty obtained by varying the resolution by $\pm 10\%$ is found 
to be even smaller. 
The estimated uncertainties are listed in Table~\ref{tab:syst}.

\begin{figure}[p]
\begin{center}
\epsfig{figure=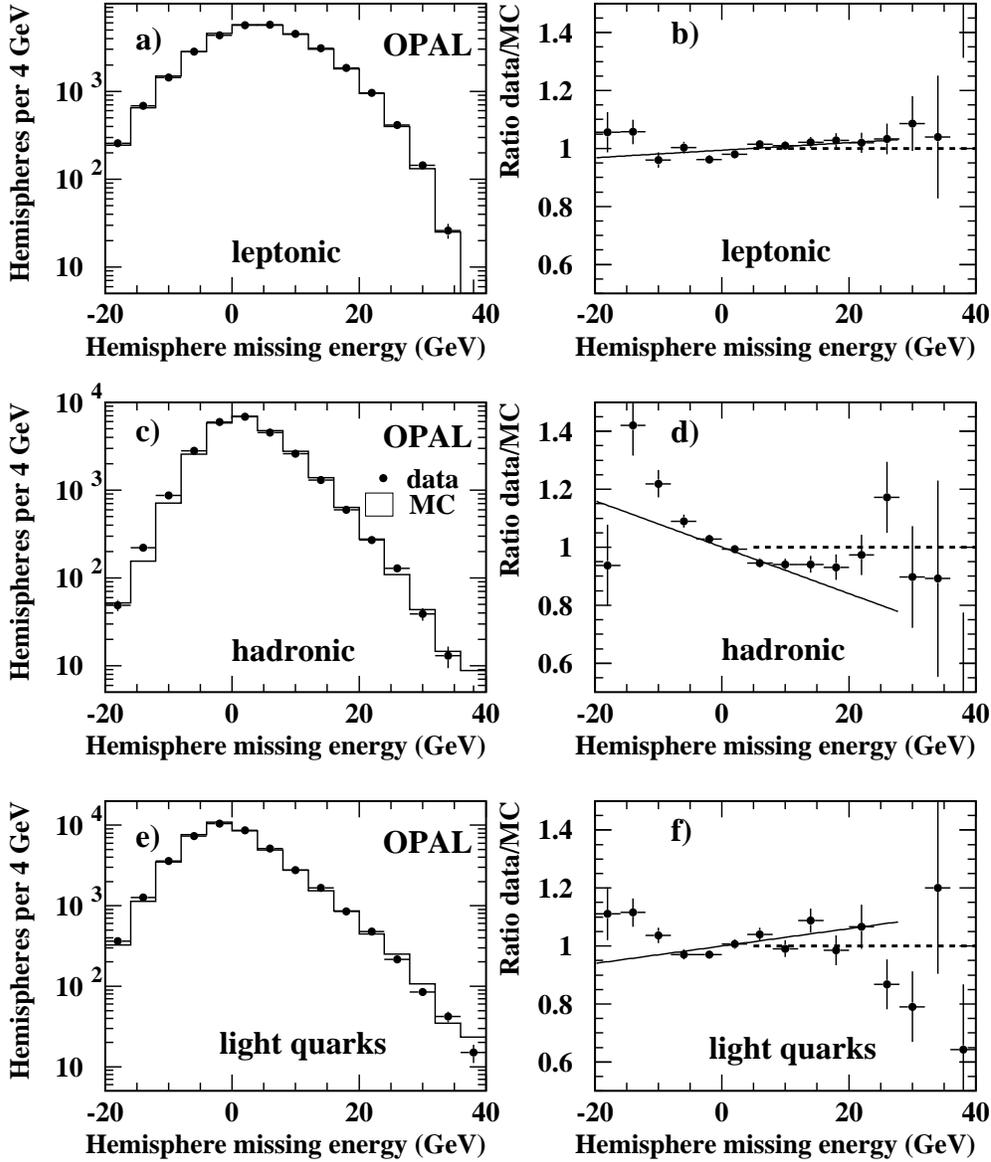,width=\textwidth}
\end{center}
\caption{Comparison of the distributions of the missing energy in a hemisphere,
$\emis$, between the data taken in 1994 and 
the Monte Carlo simulation for the three control samples:
a,b) leptonic control sample; c,d) hadronic control sample; 
e,f) light quarks control sample (see text).
The number of hemispheres in the Monte Carlo is normalised to
the number of hemispheres in the data.
The lines indicate the weight function $w(\emis)$ (Eq.~\protect\ref{eq:w})
for the fitted slopes $a$ and for $a=0$, i.e. perfect agreement
between data and Monte Carlo simulation for the control samples.
}
\label{fig:control}
\end{figure}
\end{description}

The following systematic uncertainties are related
to the modelling of the cuts:

\begin{description}

\item[b-tagging efficiency:]
The b-tagging efficiency is checked by comparing the fraction 
of events with one and two tagged hemispheres in data and Monte Carlo. 
Using the hemisphere correlation from the Monte Carlo simulation
a b-tagging efficiency of $45.7\%$ is obtained using the data compared to
$47\%$ using the Monte Carlo, i.e. the Monte Carlo is overestimating the 
b-tagging efficiency by approximately~$3\%$. 
This result is consistent with similar recent studies~\cite{bib:btag}.
The b-tagged events are reweighted according 
to this change in efficiency and the systematic shift is found to
be negligible, so no correction is applied.
In addition, the uncertainty on the rejection efficiencies for the other 
flavours is evaluated by changing the efficiencies by $10\%$.
Since the b-tagging is applied in the opposite hemisphere and
the contamination of non-b flavours in the selected sample is 
below $10\%$, the contribution to the total systematic uncertainty
is small.

\item[Lepton identification:]
An important background with missing energy due to
neutrinos originates from 
semileptonic decays of b and c quarks to muons or electrons.
In~\cite{bib:muon,bib:elec1}, the systematic uncertainties
on the efficiencies for identifying muons or electrons 
determined by the Monte Carlo are found to be $5\%$ and $4\%$.  
This uncertainty is taken into account in the efficiency for rejecting
hemispheres with electrons or muons. 
 
\end{description}

The systematic uncertainties due to the other selection cuts
are negligible. The remaining errors 
are due to fragmentation parameters and decay rates
which have to be taken from other measurements.

\begin{description}

\item[Heavy quark fragmentation and decay modelling:]
The distribution of the missing energy depends on the b 
fragmentation modelling.
The effect of the uncertainty is determined by reweighting the Monte
Carlo events to reproduce the experimental uncertainty on
the mean energy of the b hadrons, $\langle x_{\rm b}\rangle$
using Peterson fragmentation and two other 
fragmentation models~\cite{bib:colspil,bib:kartvel}.
This variation produces a different energy distribution for the b hadron decay
products, resulting in the largest systematic uncertainty.

Since only about $5\%$ of the sample is expected to
be $\Z\to\ccbar$ events, the uncertainty on $\langle x_{\rm c}\rangle$
yields only a small contribution to the total systematic uncertainty.

HQET calculations to order $1/m_\mathrm{b}^2$ show that
the $\tau$ polarisation changes by
$4\%$ compared to the free quark decay model used in the simulation.
A change in polarisation leads to a different $\tau$ energy spectrum. 
To estimate the effect of non-zero HQET parameters 
$\lambda_1$ and $\lambda_2$, the polarisation 
of the $\tau$ leptons in the Monte Carlo simulation
is varied by $4\%$ resulting in a systematic uncertainty of $0.03 \%$.
\item[Semileptonic decay models:]
A large part of the background is composed of 
semileptonic b decays where the leptons
have a momentum smaller than $2$~GeV. Both the fraction of these decays
and the shape of the missing energy distribution depend on their modelling.
The ACCMM model~\cite{bib:accmm} is used for the measurement. Using the 
ISGW~\cite{bib:isgw} model changes the result by $+0.14\%$ and using
the ISGW**~\cite{bib:isgw**} model changes the result by $-0.26\%$.
The larger of the two variations is used as
an estimate of the systematic uncertainty.
\item[Uncertainties on decay rates:]
The branching fractions of the Z boson into $\ccbar$ and $\bbbar$ pairs
in hadronic events, $R_{\rm c}$ and $R_{\rm b}$, are taken 
from~\cite{bib:pr319}. The variation of $R_{\rm c}$ and $R_{\rm b}$ within
their uncertainties yields a negligible contribution to the systematic
uncertainty.
The only significant background involving $\tau$ leptons originates from 
$\dstaunu$  decays. Averaging a recent measurement~\cite{bib:dstau}
with the measurement in~\cite{bib:dstaulep}, we obtain 
${\rm BR}(\dstaunu)=(7.2\pm2.3)\%$. 
The branching fraction of b quarks into $\dst$ is 
$f({\rm b}\to\dst)=(18\pm5)\%)$~\cite{bib:fbs}.
These uncertainties and the uncertainties on the semileptonic branching
fractions BR(\blnux) and BR(\bclnux) have 
been taken into account by reweighting the Monte Carlo events 
using the branching fractions and the uncertainties given 
in Table~\ref{tab:syst}.

\end{description}

\section{Results}
\label{sec:con}
Using about \ntot million hadronic \Z decays we have measured 
the inclusive branching ratio:
\begin{equation}
\mathrm{BR}(\btaunux) = (\brbtau \pm \brbtaustat \pm \brbtausyst) \%. 
\end{equation}
The size of the total uncertainty is similar to the uncertainties of the 
measurements in~[1-3].

A contribution from charged Higgs decays is expected to enhance the
branching ratio compared to that in the Standard Model. 
Since we have found no large enhancement of the branching 
ratio compared to the Standard Model prediction  
$\mathrm{BR}(\btaunux)=(2.36 \pm 0.17)\%$, we set a constraint
on such a contribution for Type II Two Higgs Doublet Models~\cite{bib:gross}.

The $\tau$ polarisation depends on the Higgs contribution and can be
calculated as a function of 
$r= \tan\beta / M_{{\rm H}^\pm}$~\cite{bib:gross}. 
This is taken into account by an iterative procedure 
for the limit calculation.
A limit of $r<0.52$~GeV$^{-1}$
is obtained assuming $\tau$ polarisation as for the Standard Model decay. 
The polarisation of the $\tau$ leptons in the Monte Carlo simulation
is changed according to this value of $r$ and the limit
is recalculated. The resulting limit is

\begin{equation}
\frac{\tan\beta}{M_{{\rm H}^\pm}} <  \tanblim\ {\rm GeV}^{-1}
\end{equation}
at $95\%$ confidence level. 

\section*{Acknowledgements}
We thank Y.~Grossman for helping us with the limit calculations
and P.~Urban for very useful discussions on the modelling of the 
$\btaunux$ decay.
We particularly wish to thank the SL Division for the efficient operation
of the LEP accelerator at all energies
 and for their continuing close cooperation with
our experimental group.  We thank our colleagues from CEA, DAPNIA/SPP,
CE-Saclay for their efforts over the years on the time-of-flight and
trigger
systems which we continue to use.  In addition to the support staff at our
own
institutions we are pleased to acknowledge the  \\
Department of Energy, USA, \\
National Science Foundation, USA, \\
Particle Physics and Astronomy Research Council, UK, \\
Natural Sciences and Engineering Research Council, Canada, \\
Israel Science Foundation, administered by the Israel
Academy of Science and Humanities, \\
Minerva Gesellschaft, \\
Benoziyo Center for High Energy Physics,\\
Japanese Ministry of Education, Science and Culture (the
Monbusho) and a grant under the Monbusho International
Science Research Program,\\
Japanese Society for the Promotion of Science (JSPS),\\
German Israeli Bi-national Science Foundation (GIF), \\
Bundesministerium f\"ur Bildung und Forschung, Germany, \\
National Research Council of Canada, \\
Research Corporation, USA,\\
Hungarian Foundation for Scientific Research, OTKA T-029328, 
T023793 and OTKA F-023259.\\

\clearpage

\end{document}